\documentclass[prb,twocolumn,eqsecnum,showpacs]{revtex4}

\usepackage{graphicx}

\begin{document}  

\title{Renormalization and cyclotron resonance\\
in bilayer graphene with weak electron-hole asymmetry}
\author{K. Shizuya}
\affiliation{Yukawa Institute for Theoretical Physics\\
Kyoto University,~Kyoto 606-8502,~Japan }

\begin{abstract} 

Cyclotron resonance in bilayer graphene is studied with weak electron-hole 
asymmetry, suggested by experiment, taken into account and 
with the focus on many-body corrections that evade Kohn's theorem.
It is shown by direct calculation 
that the theory remains renormalizable to $O(e^{2})$ 
in the presence of electron-hole asymmetry parameters, 
and a general program to carry out renormalization 
for graphene under a magnetic field is presented.
Inclusion of electron-hole asymmetry in part improves the theoretical fit to the existing data
and the data appear to indicate the running of the renormalized velocity factor with the magnetic field, 
which is a key consequence of renormalization.

\end{abstract} 


\maketitle

\newpage

\section{Introduction}

Graphene~\cite{NG,ZTSK,ZJS,ZA,PGN} supports charge carriers 
that behave as Dirac fermions,
which,  in a magnetic field, lead to a particle-hole symmetric 
and unequally-spaced pattern of Landau levels.
Accordingly, graphene gives rise to a variety of cyclotron resonance,
both intraband and interband, 
with resonance energies varying from one resonance to another. 
This is in sharp contrast to standard quantum Hall systems 
with a parabolic energy dispersion, where cyclotron resonance takes place 
between adjacent Landau levels, hence at a single frequency $\omega_{c} = eB/m^{*}$,
which, according to Kohn's theorem,~\cite{Kohn} 
is unaffected by electron-electron interactions.
The nonparabolic spectra~\cite{AA} in graphene offer the challenge of detecting 
many-body corrections to cyclotron resonance.

Theoretical studies~\cite{AFsr,IWFB,BMgr,KCC,KScr} over the past few years 
have revealed some notable features of
quantum corrections to cyclotron resonance in graphene 
and bilayer~\cite{NMMKF,MF} graphene.
The genuine many-body corrections arise from vacuum polarization,
specific to graphene, which diverges logarithmically at short wavelengths.
This means that one has to carry out renormalization properly, 
as in quantum electrodynamics, to extract observable results.
In particular, for bilayer graphene it turns out that 
both the leading intralayer and interlayer coupling strengths 
undergo renormalization
and that their renormalized strengths run with the magnetic field.
Bilayer graphene is marked with the unique property that its band gap 
is externally controllable.~\cite{OBSHR,Mc,CNMPL,OHL, Mucha}

Experiment has so far verified, via infrared spectroscopy, 
some basic features of cyclotron resonance in monolayer~\cite{JHT,DCN,HCJL} 
and bilayer~\cite{HJTS,OFB} graphene.
The data for the monolayer show a good symmetry 
between the electron and hole bands
but generally show no clear sign of the many-body effect,
except for a datum.~\cite{JHT}
Indeed, a comparison between 
some leading intraband and interband cyclotron resonances
revealed a small deviation in excitation energy, consistent with the presence 
of many-body corrections roughly in magnitude and sign.

The situation is quite different for bilayer graphene, 
for which only a limited number
of data are available so far.
The data~\cite{HJTS} on intraband resonances show a weak electron-hole asymmetry, 
and generally defy a good fit by theory.
Actually one has to employ different values of the velocity factor $v$ 
to fit the electron data and hole data separately.

Earlier Raman~\cite{Malard} spectroscopy and 
subsequent infrared~\cite{LHJH,ZLBF,KHML} spectroscopy  
of bilayer graphene under zero magnetic field also revealed a significant asymmetry 
between the conduction and valence bands, mainly due to subleading 
intra- and inter-layer couplings $\triangle$ and $\gamma_{4}$.

The purpose of this paper is to reexamine 
cyclotron resonance in bilayer graphene, 
with possible electron-hole and valley asymmetries taken into account.
It is not clear {\it a priori} whether the renormalizability 
of the low-energy effective theory is maintained 
in the presence of electron-hole asymmetry, 
since the asymmetry parameters (especially, $\gamma_{4}$) 
critically modify the ultraviolet structure of the theory. 
We show that the theory indeed remains renormalizable to $O(e^2)$ (at least),
and that the renormalization counterterms depend on $\gamma_{4}$ in a nontrivial way.
We present a general algorithm to carry out renormalization 
for graphene under a magnetic field, executable even numerically.
Inclusion of electron-hole asymmetry parameters partially improves 
the theoretical fit to the existing data, and the fit in turn suggests
some nontrivial modification of the spectra 
of the zero-mode and pseudo-zero-mode 
Landau levels specific to bilayer graphene.

In Sec.~II we briefly review the effective theory of bilayer graphene 
and examine the effect of electron-hole and valley asymmetries. 
In Sec.~III we study the Coulombic many-body corrections to cyclotron resonance,
with a focus on renormalization and its consequences.
Section~IV is devoted to a summary and discussion.

\section{bilayer graphene}

Bilayer graphene consists of two coupled honeycomb lattices of 
carbon atoms,
arranged in Bernal $A'B$ stacking.
The electrons in it  are described 
by four-component spinor fields on the four inequivalent sites 
$(A,B)$ and $(A',B')$ in the bottom and top layers,
and their low-energy features are governed 
by the two inequivalent Fermi points $K$ and $K'$ in the Brillouin zone. 
The intralayer coupling $\gamma_{0} \equiv \gamma_{AB} \sim 3$\, eV 
is related to the Fermi velocity 
$v = (\sqrt{3}/2)\, a_{\rm L}\gamma_{0}/\hbar \sim 10^{6}$~m/s 
(with $a_{\rm L}=  0.246$nm) in monolayer graphene.
The interlayer couplings~\cite{Malard,ZLBF} 
$\gamma_{1} \equiv \gamma_{A'B} \sim 0.4$\,eV and
$\gamma_{3} \equiv \gamma_{AB'} \sim 0.1$\,eV 
are one-order of magnitude weaker than $\gamma_{0}$.
Actually, interlayer hopping via the $(A',B)$ dimer bonds 
modifies the intralayer linear spectra
to yield quasi-parabolic spectra~\cite{MF} 
in the low-energy branches $|\epsilon| <\gamma_{1}$.

The bilayer Hamiltonian with the leading intra- and inter-layer couplings
$v \propto \gamma_{0}$ and $\gamma_{1}$ lead to 
electron-hole symmetric spectra. 
Infrared spectroscopy~\cite{ZLBF} of bilayer graphene, however, 
has detected some weak asymmetry between the electron and hole bands, 
such as (i) the energy difference $\triangle \approx 18$ meV 
between the $A$ and $B$ sublattices within the same layer and
(ii) the next-nearest-neighbor interlayer coupling 
$\gamma_{4} \equiv \gamma_{A A'} = \gamma_{B B'} \approx 0.04\, \gamma_{0}$.

The effective Hamiltonian with such intra- and inter-layer couplings 
is written as~\cite{MF,NCGP}
\begin{eqnarray}
H^{\rm bi} &=&\!\! \int\! d^{2}{\bf x}\, \Big[ \Psi^{\dag}\, {\cal H}_{+} \Psi 
+ \tilde{\Psi}^{\dag}\, {\cal H}_{-}\, \tilde{\Psi}\Big], \nonumber\\
{\cal H}_{+} &=& \left(
\begin{array}{cccc}
{1\over{2}}u &  & v_{4}\, p^{\dag}  & v\,p^{\dag} \\
  & -{1\over{2}}u & v\,p & v_{4}\, p \\
v_{4}\,p & v\,  p^{\dag} &  \triangle -{1\over{2}}u & \gamma_{1} \\
v\,p &v_{4}\, p^{\dag} & \gamma_{1}  & \triangle +{1\over{2}}u \\
\end{array}
\right),
\label{Hbilayer}
\end{eqnarray}
with $p= p_{x}+ i\, p_{y}$, $p^{\dag}= p_{x} - i\, p_{y}$.
Here $\Psi = (\psi_{A},\psi_{B'},\psi_{A'}, \psi_{B})^{\rm t}$
stands for the electron field at the $K$ valley, with $A$ and $B$ 
referring to the associated sublattices; 
$u$ stands for the interlayer bias, 
which opens a tunable gap~\cite{OBSHR} between the $K$ and $K'$ valleys.
We ignore the effect of trigonal warping 
$\propto \gamma_{3}$ 
which, in a strong magnetic field, causes only a negligibly small level shift.~\cite{KSbgr}
We also ignore weak Zeeman coupling and, for conciseness, 
suppress the electron spin.
Our definition of $v_{4} \equiv -(\gamma_{4}/\gamma_{0})\, v$ differs 
in sign and by factor $v$ from the one ($v_{4}\rightarrow \gamma_{4}/\gamma_{0}$) 
in the literature~\cite{Malard,LHJH,ZLBF,NCGP};
this choice is made simply for notational convenience.

The Hamiltonian ${\cal H}_{-}$ at another ($K'$) valley is given by ${\cal H}_{+}$ 
with $(v, v_{4},u) \rightarrow (-v, -v_{4}, -u)$, 
and acts on a spinor of the form 
$\tilde{\Psi} = (\psi_{B'},\psi_{A},\psi_{B}, \psi_{A'})^{\rm t}$.
Note that ${\cal H}_{+}$ is unitarily equivalent to 
${\cal H}_{-}$ with the sign of $u$ reversed,  
\begin{equation}
U^{\dag}{\cal H}_{+}|_{u}U ={\cal H}_{-}|_{-u}
\label{Hequi}
\end{equation}
with $U= {\rm diag}(1,1,-1,-1)$.
This implies that the electronic spectrum at the $K'$ valley 
is obtained from the spectrum at the $K$ valley
by reversing the sign of $u$; 
in particular, the spectra at the two valleys are the same for $u=0$.
Nonzero interlayer voltage $u\not=0$ thus acts as a valley-symmetry breaking.

We adopt the set of experimental values~\cite{ZLBF}
\begin{eqnarray}
v &\approx&1.1 \times 10^{6}\, {\rm m/s},\  
\gamma_{1} \approx 404\, {\rm meV}, \nonumber\\
v_{4}/v &\equiv& -\gamma_{4}/\gamma_{0}\approx - 0.04,\ 
\triangle \approx  18\, {\rm meV}, 
\label{expparameters}
\end{eqnarray}
in what follows.  Full account is also taken of the effect of interlayer bias $u$. 
For notational simplicity, however,  we often present analytical expressions 
only for $u=0$.

The Hamiltonian $H^{\rm bi}$ gives rise to four bands with electron and hole spectra,
which, for $u=0$, read
\begin{eqnarray}
\epsilon_{4}({\bf p}) &=&\sqrt{ v_{+}^{2}\, {\bf p}^{2} + (\gamma_{+}/2)^{2}} 
+ \gamma_{+}/2, \nonumber\\
\epsilon_{3}({\bf p}) &=&\sqrt{ v_{-}^{2}\, {\bf p}^{2} + (\gamma_{-}/2)^{2}} 
- \gamma_{-}/2, \nonumber\\
\epsilon_{2}({\bf p}) &=&-\sqrt{ v_{+}^{2}\, {\bf p}^{2} + (\gamma_{+}/2)^{2}} 
+ \gamma_{+}/2, \nonumber\\
\epsilon_{1}({\bf p}) &=& -\sqrt{ v_{-}^{2}\, {\bf p}^{2} + (\gamma_{-}/2)^{2}} 
- \gamma_{-}/2, 
\label{Eip}
\end{eqnarray}
where $v_{\pm} \equiv v \pm v_{4}$ 
and $\gamma_{\pm} \equiv \gamma_{1} \pm \triangle$.
Note that $v_{4}$ and  $\triangle$ effectively modify $v$ and $\gamma_{1}$, respectively,
in a manner different for electrons and holes; 
the spectra are electron-hole asymmetric unless $v_{4}=\triangle=0$.
These band spectra acquire nonzero valley gaps for $u\not=0$.

Let us place bilayer graphene in a strong uniform magnetic field 
$B_{z} = -B<0$ normal to the sample plane;
we set, in ${\cal H}_{\pm}$, 
$p\rightarrow \Pi = p + eA$ and $p^{\dag}\rightarrow \Pi^{\dag}$ 
with $A= A_{x}+ iA_{y}= B\, y$, 
and denote the the magnetic length as $\ell=1/\sqrt{eB}$.
It is easily seen that the eigenmodes of ${\cal H}_{+}$ have the structure 
\begin{equation}
\Psi_{n} = \Big(|n\rangle\, b_{n}^{(1)} ,|n\!-\!2\rangle\, b_{n}^{(2)},
|n\!-\!1\rangle\, b_{n}^{(3)}, |n\!-\!1\rangle\, b_{n}^{(4)}\Big)^{\rm t}
\end{equation} 
with $n=0,1,2,\dots$, where only the orbital eigenmodes are shown 
using the standard harmonic-oscillator basis $\{ |n\rangle \}$
(with the understanding that $|n\rangle =0$ for $n<0$).
The coefficients 
${\bf b}_{n}=(b_{n}^{(1)}, b_{n}^{(2)}, b_{n}^{(3)}, b_{n}^{(4)})^{\rm t}$
 for $n=2,3,\dots$ are given by the eigenvectors of the reduced Hamiltonian
\begin{equation}
\hat{\cal H}_{\rm red} = \omega_{c}\, \left(
\begin{array}{cccc}
{1\over{2}}u' &  \ & r \sqrt{n}  & \sqrt{n} \\
 & - {1\over{2}}u' & \sqrt{n\!-\!1} & r \sqrt{n\!-\!1} \\
r \sqrt{n} & \sqrt{n\!-\!1} \ & d - {1\over{2}}u' & \gamma' \\
\sqrt{n} & r \sqrt{n\!- \! 1} & \gamma'  & d + {1\over{2}}u' \\
\end{array}
\right),
\label{reducedH}
\end{equation} 
where 
\begin{equation}
\omega_{c}\equiv \sqrt{2}\, v/\ell 
\approx 36.3 \times v[10^{6}{\rm m/s}]\, \sqrt{B[{\rm T}]}\ {\rm meV},
\end{equation} 
with $v$ measured in units of $10^{6}$m/s and $B$ in tesla, 
is the characteristic cyclotron energy 
for monolayer graphene;
$r \equiv  v_{4}/v = - \gamma_{4}/\gamma_{0}\ (\approx -0.04)$, 
$\gamma' \equiv  \gamma_{1}/\omega_{c}$, 
$d \equiv  \triangle/\omega_{c}$ and $u'\equiv u/\omega_{c}$.

The energy eigenvalues $\epsilon_{n}$ of $\hat{\cal H}_{\rm red}$ 
are determined from the secular equation, which, for $u=0$, reads
\begin{eqnarray}
&& n (n-1) (1-r^{2})^{2} 
- \Big[\gamma'^{2} -d^{2} + c_{n}\, (1\!+ r^{2}) \Big]\, \epsilon'^{2} 
+ (\epsilon'^{2})^{2}
\nonumber\\
&& -c_{n}\, [2\, r\, \gamma' - d\, (1+r^{2})]\, \epsilon'
-2d\, (\epsilon')^{3}=0,
\label{seceq}
\end{eqnarray}
with $c_{n}= 2n -1$ and $\epsilon'\equiv \epsilon_{n}/\omega_{c}$.
We first consider the $u=0$ case. 
Let us  denote the four solutions of the secular equation as
$\epsilon_{-n}^{-} < \epsilon_{-n}<0<\epsilon_{n}
< \epsilon_{n}^{+}$ for each integer $n\ge 2$, 
so that the index $\pm n$ reflects the sign of the energy eigenvalues. 
For $n=0$ $H^{\rm bi}$ has an obvious zero eigenvalue 
$\epsilon_{0} =0$, with the eigenvector
${\bf b}_{0} = (1,0,0,0)^{\rm t}$.
For $n=1$ the secular equation~(\ref{seceq}) is reduced 
to a cubic equation in $\epsilon'$,
excluding $\epsilon'=0$, and leads to three solutions,
which, for the present choice~(\ref{expparameters}) of parameters,
are given, e.g., by $\epsilon' =(-3.96, 0.029, 4.29)$ at $B=10$ T; 
we thus denote the corresponding eigenvalues
as $(\epsilon^{-}_{-1}, \epsilon_{1},\epsilon^{+}_{1})$.
This eigenvalue $\epsilon_{1}$ changes sign if one sets $d\rightarrow -d$ and
$r \rightarrow -r$ simultaneously. Thus the assignment of $\epsilon_{\pm1}$
in general depends on the choice of asymmetry parameters $(v_{4}, \triangle)$ 
and also on the interlayer bias $u$.
Actually, for zero bias $u=0$, $\epsilon_{1}$ deviates from zero 
as $v_{4}$ and $\triangle$ develop.
In this sense, the $n= 1$ Landau level is a pseudo-zero-mode level
while the $n=0$ level is a genuine zero-mode level.

As $u$ is turned on, these $n= (0, \pm 1)$ levels go up or down oppositely 
at the two valleys; e.g., $\epsilon_{0}|_{K} = - \epsilon_{0}|_{K'} = u/2$.
Their spectra vary linearly with $u$ while other levels $(|n| \ge 2)$ 
get shifted only slightly.
Interestingly, for $0<|u|  \ll \omega_{c}$, $\epsilon_{1} \gtrsim \epsilon_{0_{+}}$ 
at one valley 
while $\epsilon_{|n|=1}$ at another valley crosses~\cite{fnx} $\epsilon_{0_{+}}$
from below with increasing magnetic field $B$.
A nonzero $u$ thus  critically spoils the valley symmetry of the $n= (0,\pm 1)$ sector.

The spectra $\epsilon^{+}_{n}$ and $\epsilon^{-}_{-n}$ with $n\ge1$
form the high-energy branches of the electron and hole Landau levels, respectively;
$|\epsilon^{\pm}|\gtrsim \gamma_{1}$.
Let us combine $\epsilon_{\pm n}$ into the low-energy branch of Landau levels
$\{\epsilon_{n}\}
= \{ \dots,\epsilon_{-3},\epsilon_{-2}, 
\epsilon_{0}, \epsilon_{1}, \epsilon_{2}, \dots\}$,
and denote the three branches 
$(\epsilon_{-n}^{-}, \epsilon_{n}, \epsilon_{n}^{+})$ 
as
\begin{equation}
\epsilon_{n}= \omega_{c}\, \eta_{n}(\gamma', d, r, u'), \ 
\epsilon^{\pm}_{n}= \omega_{c}\, \eta^{\pm}_{n}(\gamma', d, r, u') .
\label{En}
\end{equation} 
These $(\eta_{n},\eta^{\pm}_{n}) \sim \epsilon'$ are uniquely determined 
from $\hat{\cal H}_{\rm red}$ of Eq.~(\ref{reducedH}) 
or from Eq.~(\ref{seceq}) as functions of $(\gamma', d, r, u')$.
One can thereby construct 
the associated eigenvectors, which, e.g., 
for $u=0$ and $|n|\ge 1$, read
${\bf b}_{n}=(b_{n}^{(1)}, b_{n}^{(2)}, b_{n}^{(3)}, b_{n}^{(4)})^{\rm t} 
= b_{n}^{(1)}\, ( 1, \beta_{n}^{(2)}, \beta_{n}^{(3)}, \beta_{n}^{(4)})^{\rm t}$ 
with
\begin{eqnarray}
\beta_{n}^{(3)} &=& -{|n| -\eta_{n}\, (\eta_{n} -d)
 + (|n|-1)\, r^{2}
\over{\sqrt{|n|}\, G_{n}}},
\nonumber\\
\beta_{n}^{(4)} &=& \Big[\gamma'\, \eta_{n} + (2|n|-1)\, r \Big]
/ \Big(\sqrt{|n|}\, G_{n}\Big),
\nonumber\\
\beta_{n}^{(2)} &=&  \sqrt{|n|-1}\,( \beta_{n}^{(3)} +r\, \beta_{n}^{(4)})/\eta_{n},
\end{eqnarray}
where  $G_{n} = \gamma' + r\, \{\eta_{n} -d +  (|n|-1)\, (1-r^{2})/\eta_{n} \}$
and $b_{n}^{(1)}= 1/\sqrt{1 +\sum_{i=2,3,4}(\beta_{n}^{(i)})^{2}}$.
These expressions are equally valid for eigenvectors belonging to $\epsilon_{n}^{\pm}$,
with $\eta_{n} \rightarrow \eta_{n}^{\pm}$.


\begin{figure}[tbp]
\includegraphics[scale=0.25]{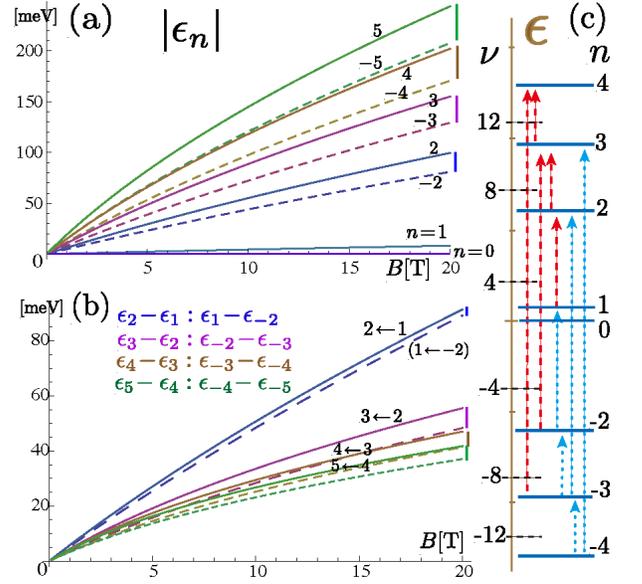}  
\caption{The $3 \leftarrow -2$ (or $2 \leftarrow -3$) resonance
(a) Landau levels 
as a function of magnetic field $B$
for $\gamma_{1}=404$ meV, $\triangle= 18$ meV, $v_{4}/v = -0.04$ and $u=0$;  
$\epsilon_{n}$ for electrons (solid curves) and 
$-\epsilon_{-n}$ for holes (dashed curves). 
(b) Landau  gaps for electrons (solid curves)  
and holes (dashed curves).
(c) Typical channels of cyclotron resonance for $u=0$;
circularly polarized light can distinguish between the two classes of transitions
indicated by different types of arrows.
}
\end{figure}

The Landau-level spectrum 
$\epsilon_{n} = \omega_{c}\, \eta_{n}(\gamma', d, r, u')$ 
depends on the magnetic field $B$ in a nontrivial manner
through the dimensionless quantities $\gamma'$, $d$ and $u'$.
Actually, for the choice of $\gamma_{1} =404$ meV, 
$\triangle =18$ meV,  $r= -0.04$ and $u=0$, 
the electron and hole spectra differ considerably, as shown in Fig.~1.
In particular, the zero-mode level $(n=0)$ remains intact 
while the pseudo-zero-mode level ($n=1$) gets shifted, e.g., 
by $\sim$ 8 meV as $B$ is increased from 0 to 20~T.
The Landau gaps are generally larger for electrons than holes;
this is readily understood from the behavior of 
$\epsilon_{3}({\bf p}) \approx (v_{-}^{2}/\gamma'_{-})\, {\bf p}^{2}$
for ${\bf p}\rightarrow 0$, which implies 
that the effective mass $m^{*} \sim \gamma_{-}/(2\, v_{-}^{2})$
is smaller for electrons. 
The asymmetry between the electron Landau levels $L_{n}$ and 
hole levels $L_{-n}$ becomes more prominent for higher levels $|n|\ge 2$.


\begin{figure}[tbp]
\includegraphics[scale=0.195]{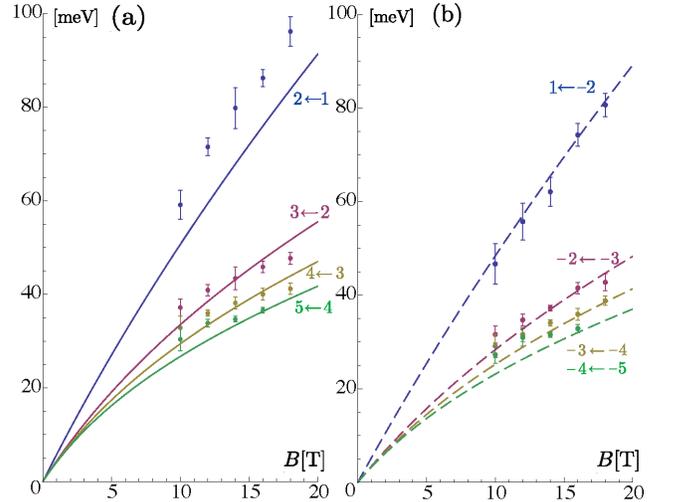}  
\caption{
Landau gaps in Fig.~1~(b), compared with the experimental data
in Ref.~\onlinecite{HJTS} on cyclotron resonance energies.
(a)~Electron band.  (b)~Hole band.
}
\end{figure}

Cyclotron resonance in bilayer graphene is governed 
by the selection rule~\cite{AFsr} $\triangle |n| =\pm 1$, 
and there are two classes of transitions, (i) $L_{\pm (n-1)} \leftarrow L_{-n}$ and 
(ii) $L_{n} \leftarrow L_{\pm (n-1)}$ (with $n\ge 2$), which are distinguished 
by the use of circularly-polarized light. 
See Fig.~1~(c).

As for experiment, Henriksen {\it et al.}~\cite{HJTS} measured, 
via infrared spectroscopy, 
cyclotron resonance in bilayer graphene in magnetic fields up to 18T. 
They observed intraband resonances, which are identified with
the $L_{2}\leftarrow  L_{1}$, $L_{3}\leftarrow  L_{2}$, $L_{4}\leftarrow  L_{3}$ 
and $L_{5}\leftarrow  L_{4}$ transitions
at filling factor $\nu$= 4, 8, 12 and 16, respectively, 
and the corresponding hole resonances at $\nu$= -4, -8, -12 and -16,
together with a significant asymmetry between the electron and hole data.

Such data from Ref.~\onlinecite{HJTS} are reproduced in Fig.~2.
Also included are the zeroth-order Landau gaps of Fig.~1~(b), 
which apparently fit the experimental data reasonably well,
except for the $\nu=4$ electron data on the $L_{2}\leftarrow L_{1}$ resonance, 
which deviates considerably.

It is worth discussing the effect of interlayer bias $u$ here.
The $n=(0,1)$ levels are very sensitive to $u$
and may easily acquire valley gaps for $u\not=0$ while 
other levels $|n|\ge 2$ are relatively inert as long as $|u|\ll \omega_{c}$.
Cyclotron resonances involving the $n=1$ level, 
i.e., $(1\leftarrow -2)$ and $(2\leftarrow 1)$ resonances,
therefore tend to be affected by $u$.
Actually, with $u\sim 20\, {\rm meV}$ 
one can apparently fit the data for those resonances
at one valley.
The asymmetry, however, is reversed at another valley. 
Nonzero $u$ may thus broaden the observed widths 
of the $(1\leftarrow -2)$ and $(2\leftarrow 1)$ resonances 
but would not account for their asymmetry. 

It is clear now that one should treat those resonances separately 
from the rest of the resonances, which are barely sensitive to $|u|\ll \omega_{c}$.
The former and latter are also different in their sensitivities
to electron-hole asymmetry $\propto \triangle$ and $v_{4}$.
See Fig.~1~(b) again. It shows 
that for $u=0$ the $2\leftarrow 1$ and $1\leftarrow -2$ gaps barely differ
while other Landau gaps exhibit significant asymmetry 
between the electron and hole bands.
In this sense, Fig.~2 shows us that 
$\triangle \approx 18$meV and $v_{4}\approx -0.04\, v$,
obtained from independent experiments, account for the electron-hole asymmetry 
between the $\nu=(8,12,16)$ data and the $\nu= (-8,-12, -16)$ data reasonably well.
It is a nontrivial fact that this single set of parameters
can fit the electron and hole data simultaneously.

Poor fitting to the $\nu= 4$ data, on the other hand, 
would suggest that the spectrum of the pseudo-zero-mode level $(|n|=1)$
is further modified~\cite{AbCh}  by some other sources. 
Actually it is expected theoretically~\cite{BCNM,KSpzm,CLBM} 
that the $n=(0,\pm 1)$ sector of bilayer graphene 
has nontrivial dynamics due to orbital mixing 
and supports characteristic collective excitations, orbital pseudospin waves. 
It would be interesting to study how the electron-hole asymmetry 
affects the detailed structure of this special sector.

\section{Cyclotron resonance and many-body corrections}

In this section we study the many-body corrections to cyclotron resonance, 
with emphasis on how to carry out renormalization.
The Coulomb interaction is written as
\begin{equation}
H^{\rm Coul} 
= {1\over{2}} \sum_{\bf p}
v_{\bf p}\, :\rho_{\bf -p}\, \rho_{\bf p}:,
\label{Hcoul}
\end{equation}
where $\rho_{\bf p}$ is the Fourier transform of the electron
density $\rho = \Psi^{\dag}\Psi + \tilde{\Psi}^{\dag}\tilde{\Psi}$;
$v_{\bf p}= 2\pi \alpha/(\epsilon_{\rm b} |{\bf p}|)$ is 
the Coulomb potential with 
$\alpha = e^{2}/(4 \pi \epsilon_{0}) \approx 1/137$ and 
the substrate dielectric constant $\epsilon_{\rm b}$;
$\sum_{\bf p} =\int d^{2}{\bf p}/(2\pi)^{2}$.

The Landau-level structure is made explicit by passing to
the $|n,y_{0}\rangle$ basis  (with $y_{0}\equiv \ell^{2}p_{x}$) 
via the expansion
$(\Psi ({\bf x}, t), \tilde{\Psi} ({\bf x}, t) ) 
= \sum_{n, y_{0}} \langle {\bf x}| n, y_{0}\rangle\, \psi_{n}(y_{0},t)$; 
remember that fields $\psi_{n}$ carry (suppressed) spin and valley indices.
The Hamiltonian $H^{\rm bi}$ is thereby rewritten as
\begin{equation}
H^{\rm bi} =  \int\! dy_{0} \!\!\!
\sum_{n =-\infty}^{\infty} \!\!\!
\psi^{\dag}_{n} (y_{0},t)\, \epsilon_{n}\, \psi_{n} (y_{0},t),
\label{Hzeronn}
\end{equation}
and the charge density $\rho_{-{\bf p}}(t) =\int d^{2}{\bf x}\,  
e^{i {\bf p\cdot x}}\,\rho$ as~\cite{KSbgr}
\begin{eqnarray}
\rho_{-{\bf p}} &=&\sum_{k, n=-\infty}^{\infty} \rho^{k n}_{\bf -p}
=\sum_{k, n=-\infty}^{\infty} g^{k n}_{\bf p}\, 
R^{k n}_{\bf p}, \nonumber\\
R^{kn}_{\bf p}&=& \gamma_{\bf p}\int dy_{0}\,
\psi_{k}^{\dag}(y_{0},t)\, e^{i{\bf p\cdot r}}\,
\psi_{n} (y_{0},t),
\label{chargeoperator}
\end{eqnarray}
where $\gamma_{\bf p} =  e^{- \ell^{2} {\bf p}^{2}/4}$; 
${\bf r} = (i\ell^{2}\partial/\partial y_{0}, y_{0})$
stands for the center coordinate with uncertainty 
$[r_{x}, r_{y}] =i\ell^{2}$.
The charge operators $R^{k n}_{\bf p}$ obey 
two $W_{\infty}$ algebras~\cite{GMP}
associated with intralevel center-motion 
and interlevel mixing of electrons.

The coefficient matrix $g^{kn}_{\bf p}$ is constructed 
from the knowledge of the eigenvectors ${\bf b}_{n}$,
\begin{eqnarray}
g^{kn}_{\bf p} &=& b_{k}^{(1)}\, b_{n}^{(1)}\, f_{\bf p}^{|k|,|n|}
+ b_{k}^{(2)}\, b_{n}^{(2)}\, f_{\bf p}^{|k|-2,|n|-2} \nonumber\\
&&+ (b_{k}^{(3)}\, b_{n}^{(3)}+ b_{k}^{(4)}\, b_{n}^{(4)})\, f_{\bf p}^{|k|-1,|n|-1},
\label{gkn}
\end{eqnarray}
where
\begin{equation}
f^{k n}_{\bf p} 
= \sqrt{{n!\over{k!}}}\,
\Big({-\ell p\over{\sqrt{2}}}\Big)^{k-n}\, L^{(k-n)}_{n}
\Big ({1\over{2}} \ell^{2}{\bf p}^{2}\Big)
\label{fknp}
\end{equation}
for $k \ge n\ge0$, and $f^{n k}_{\bf p} = (f^{k n}_{\bf -p})^{\dag}$;
$p=p_{x}\! +i\, p_{y}$. 
Expression~$(\ref{gkn})$ is valid for $|n|$ = 0, 1 as well,
with the understanding that $f^{kn}_{\bf p}=0$ for $k<0$ or $n<0$.

For zero bias $u=0$, $g^{k n}_{\bf p}$ 
are the same at the two valleys, 
i.e., $g^{k n}_{\bf p}|_{K'}= g^{k n}_{\bf p}|_{K}$
with 
$(b_{n}^{(1)}, b_{n}^{(2)}, b_{n}^{(3)}, b_{n}^{(4)})|_{K'}
= (b_{n}^{(1)}, b_{n}^{(2)}, -b_{n}^{(3)}, -b_{n}^{(4)})|_{K}$.
This follows from the unitary equivalence~(\ref{Hequi}) 
of the Hamiltonians ${\cal H}_{\pm}|_{\pm u}$ 
and the invariance of the charge density $\rho_{\bf p}$ under
$U$ there.

The Coulombic correction to cyclotron resonance in graphene 
to $O(\alpha/\ell)$ was calculated earlier~\cite{KScr} 
using the single-mode approximation.~\cite{GMP}
Here we consider cyclotron resonance (at integer filling $\nu$)  
from the filled $a$th Landau level $(L_{a})$ 
to the empty $b$th level $(L_{b})$
at zero momentum transfer ${\bf k}=0$,
where no mixing takes place in spin and valley.
The cyclotron-resonance energy 
for a general $L_{b}\leftarrow L_{a}$ transition 
with the Landau levels filled up to $n=n_{\rm f}$ $(a\le n_{\rm f}<b)$
is written as~\cite{KScr}
\begin{equation}
\epsilon^{b\leftarrow a}_{\rm exc} = \epsilon_{b} -\epsilon_{a} 
+ \triangle \epsilon^{b\leftarrow a} ,
\label{Eexc}
\end{equation}
with the correction 
\begin{equation}
\triangle \epsilon^{b\leftarrow a} = \sum_{\bf p}v_{\bf p}\,
\gamma_{\bf p}^{2}\, \Big[
\sum_{n\le n_{\rm f}} (|g^{an}_{\bf -p}|^{2}- |g^{b n}_{\bf p}|^{2})
- g^{bb}_{\bf p}\, g^{aa}_{\bf -p}\Big] ,
\nonumber\\
\label{dEba}
\end{equation}
diagonal in spin and valley.
As shown by Fig.~1~(c), $n_{\rm f}=-4,-3,-2,1,2,3, ...$ 
correspond to the filling factor
$\nu = -12, -8, -4,4,8,12,...$,
respectively.

One can now substitute Eq.~(\ref{gkn}) into this formula 
and calculate the Coulombic corrections
with the effect of electron-hole asymmetry taken into account.
There is, however, one technical problem to solve. 
The $\sum_{n\le n_{\rm f}} (|g^{an}_{\bf p}|^{2}- |g^{bn}_{\bf -p}|^{2})$ term
in Eq.~(\ref{dEba}) refers to quantum fluctuations of the filled states 
and actually diverges logarithmically 
with the number  $N_{\rm L} \rightarrow \infty$ of filled Landau levels 
in the valence band (or the Dirac sea).

One has to handle such ultraviolet (UV)
divergences by renormalization of the basic parameters 
$(v, v_{4}, \gamma_{1}, \triangle)$ and, if necessary, 
the tunable parameter $u$.
In the electron-hole symmetric case, $v$ and $\gamma_{1}$ 
turn out to be renormalized 
in the same way,~\cite{KScr} i.e., $v= Z_{v}\, v^{\rm ren}$ and 
$\gamma_{1}= Z_{\gamma}\, \gamma_{1}^{\rm ren}$ with
$Z_{v}=Z_{\gamma}$ to $O(\alpha/\ell)$.
Actually, it is not clear {\it a priori} if the theory remains renormalizable
in the presence of asymmetry parameters $v_{4}$ and $\triangle$.
If inclusion of $v_{4}$ and $\triangle$ were to yield a new type of divergence
unremovable by rescaling of the existing parameters, 
the theory would lose renormalizability 
(or one would have to introduce a new parameter to remove the divergence
and, if necessary, repeat this process).
We prove by direct calculations below that 
the theory is renormalizable to $O(\alpha/\ell)$.

The key to this problem of renormalization is 
to note that the magnetic field supplies 
only a long-wavelength cutoff through the magnetic length $\ell = 1/\sqrt{eB}$, 
leaving the UV structure of the theory intact.
One can therefore first  look into the theory in free space $(B=0)$ and 
determine the UV structure of
the Coulomb exchange corrections.
Such corrections are written as $\sum_{\bf k}v_{\bf k}\, iS({\bf p+k})$,
a convolution of the photon propagator 
$v_{\bf k} =2\pi \alpha/(\epsilon_{b} |{\bf k}|)$ and 
the instantaneous electron propagator
$\langle \Psi \Psi^{\dag}\rangle_{t=t'}\stackrel{\rm F.T.}{=}iS({\bf p})$.
Their UV structure is thus read from the asymptotic behavior of $S({\bf p})$.

The resulting divergences are then absorbed into the counterterms 
$\delta v$, $\delta \gamma_{1}$,
$\delta v_{4}$, $\delta \triangle$ and $\delta u$, generated by rescaling 
\begin{eqnarray}
v&=&Z_{v}\, v^{\rm ren}= v^{\rm ren} + \delta v,  \nonumber\\
\gamma_{1}&=&\gamma_{1}^{\rm ren} + \delta \gamma_{1}, \dots,
\end{eqnarray}
where $\lq\lq$ren" refers to renormalized parameters.
See Appendix A for such an analysis of divergences.
Here we quote only the result: 
(i)~The counterterm for velocity factor $v$ turns out to be the same as 
in the electron-hole symmetric case 
(and in the case of monolayer graphene~\cite{velrenorm} as well),
\begin{equation}
\delta v = (Z_{v} -1)\, v^{\rm ren} \sim 
 -(\alpha/8 \epsilon_{b})\, \log \Lambda^{2},   
\label{deltav}
\end{equation}
where $\Lambda$ stands for the momentum cutoff 
which is related to the Dirac-sea cutoff 
so that~\cite{KScr} $\Lambda^{2}\approx 2N_{L}/\ell^{2}$. 
(ii)~Remarkably, $v_{4}$ and $u$ remain finite,
\begin{equation}
\delta v_{4}= \delta u = 0,
\end{equation} 
and require no renormalization, $v_{4}=v_{4}^{\rm ren}$ and $u = u^{\rm ren}$.
(iii)~The dimensional parameters $\gamma_{1}$ and $\triangle$ 
are mixed under renormalization,
\begin{eqnarray}
\delta \gamma_{1}&=& (\gamma_{1}^{\rm ren}- r^{\rm ren}\, \triangle^{\rm ren})\, 
h[r^{\rm ren}]\, \delta v/v^{\rm ren},  \nonumber\\
\delta \triangle &=& 2\, (\triangle^{\rm ren}- r^{\rm ren}\, \gamma_{1}^{\rm ren})\,
h[r^{\rm ren}]\, \delta v/v^{\rm ren},
\label{deltagamma}
\end{eqnarray}
where $h[r]= 1/(1-r^{2})$ and $r^{\rm ren} = v_{4}^{\rm ren}/v^{\rm ren}$.
Note that the counterterms are highly nonlinear
in $r^{\rm ren} \propto v_{4}$.

One can now pass to the $B\not=0$ case with these counterterms.
Let us denote by ${\cal H}_{+}^{\rm ren}$ the Hamiltonian ${\cal H}_{+}$
[of Eq.~(\ref{Hbilayer})] in magnetic field $B$
with $(v,v_{4},\gamma_{1}, \triangle, u)$ replaced by 
$(v^{\rm ren}, v_{4}^{\rm ren}, \gamma_{1}^{\rm ren}, 
\triangle^{\rm ren}, u^{\rm ren})$, 
and write its spectrum 
as $\epsilon_{n}^{\rm ren} = \omega_{c}^{\rm ren}\, 
\eta_{n}(\gamma'^{\rm ren}, d^{\rm ren}, r^{\rm ren}, {u'}^{\rm ren})$ with
$\omega_{c}^{\rm ren} = \sqrt{2}\, v^{\rm ren}/\ell$, etc.,
in obvious notation; see Eq.$~(\ref{En})$.
Suppose now that we start with ${\cal H}_{+}^{\rm ren}$
and calculate Coulombic corrections to $O(\alpha/\ell)$.
The divergences we encounter are removed by the counterterms 
formally written as
$\delta_{\rm ct}{\cal H}_{+}^{\rm ren}$, where the differential operator
\begin{equation}
\delta_{\rm ct} = \delta v{\partial\over{\partial v^{\rm ren}}} 
+ \delta \gamma_{1} {\partial\over{\partial \gamma_{1}^{\rm ren}}} 
+\delta \triangle {\partial\over{\partial\triangle^{\rm ren}}}  
\end{equation}
acts on ${\cal H}_{+}^{\rm ren}$.
For the related reduced Hamiltonian 
$\hat{\cal H}_{\rm red}^{\rm ren}$, 
defined as in Eq.~(\ref{reducedH}),
the counterterm is also written as 
$\delta_{\rm ct}\hat{\cal H}_{\rm red}^{\rm ren}$.
If, for example, one is to subtract divergences from the $O(\alpha/\ell)$ correction 
to the spectrum $\epsilon_{n}^{\rm ren}$, the required counterterm 
is obtained from the expectation value of 
$\delta_{\rm ct}\hat{\cal H}_{\rm red}^{\rm ren}$, 
which equals~\cite{fndH} $\delta_{\rm ct}\, \epsilon_{n}^{\rm ren}$, 
the variation of the eigenvalue itself.
One can equally handle it numerically by writing
\begin{equation}
\delta_{\rm ct}\, \epsilon_{n}^{\rm ren} = ({\bf b}_{n})^{\dag}\! \cdot 
\delta_{\rm ct}\hat{\cal H}_{\rm red}^{\rm ren}\! 
\cdot  {\bf b}_{n}.
\label{matrixct}
\end{equation} 
Rewriting $\delta_{\rm ct}$ in favor of renormalized parameters,
$\delta_{\rm ct}
= (\delta v/v^{\rm ren} )\,( v^{\rm ren} \partial/\partial v^{\rm ren}
+ {\cal D} )$, yields
\begin{eqnarray}
\delta_{\rm ct} \epsilon_{n}^{\rm ren} &=& \omega_{c}^{\rm ren}\, 
(\delta v/v^{\rm ren})\
(\eta_{n} + {\cal D}\, \eta_{n} ), \nonumber\\
{\cal D}&=& - r\,{\partial\over{\partial r}}  
+ {r (r\, \gamma'- d)\over{1-r^{2}}}\, {\partial\over{\partial \gamma'}} \nonumber\\
&&+ { (1+ r^{2})\, d - 2\, r\, \gamma'\over{1-r^{2}}}\,{\partial\over{\partial d}}\ ; 
\end{eqnarray}
for conciseness, we have suppressed $"$ren" in ${\cal D}$.

With this in mind, let us rewrite Eq.~(\ref{Eexc}) as
\begin{equation}
\epsilon_{\rm exc}^{b \leftarrow a} = \epsilon_{b}^{\rm ren} -\epsilon_{a}^{\rm ren} 
+ (\triangle \epsilon^{b a})^{\rm ren},
\end{equation}
where the renormalized correction $(\triangle \epsilon^{ba})^{\rm ren} 
\equiv \delta_{\rm ct}\epsilon_{b}^{\rm ren}- \delta_{\rm ct}\epsilon_{a}^{\rm ren} 
+ \triangle \epsilon^{ba}$ is now made finite.
Writing the counterterm as
$\delta_{\rm ct}\epsilon_{b}^{\rm ren}- \delta_{\rm ct}\epsilon_{a}^{\rm ren} 
=\eta^{b\leftarrow a} \, \omega_{c}^{\rm ren}\, (\delta v/v^{\rm ren})$, 
with
\begin{equation}
\eta^{b\leftarrow a} \equiv \eta_{b}- \eta_{a} + {\cal D}(\eta_{b}- \eta_{a}),
\label{etaba}
\end{equation}
and setting $\triangle \epsilon^{b\leftarrow a}/ \eta^{b\leftarrow a}
\equiv V_{c}\, c^{b\leftarrow a}$, in units of the characteristic Coulomb energy
\begin{equation}
V_{c} \equiv \alpha/(\epsilon_{b}\ell) \approx (56.1/\epsilon_{b})\, 
\sqrt{B[{\rm T}]}\, {\rm meV},
\end{equation}
yields the expression
\begin{equation}
(\triangle \epsilon^{ba})^{\rm ren} 
=\eta^{b\leftarrow a} \, \{ V_{c}\, c^{b\leftarrow a} + (\sqrt{2}/\ell)\, \delta v \}.
\label{dEscale}
\end{equation}
This reveals that 
the UV divergence is common to all ratios 
$\triangle \epsilon^{b\leftarrow a}/\eta^{b\leftarrow a} = V_{c}\, c^{b\leftarrow a}$,
independent of $(b, a)$ and $(v,v_{4},\gamma_{1}, \triangle)$;
the dimensionless quantities $c^{b\leftarrow a}$ 
have the structure $c^{b\leftarrow a} = (\sqrt{2}/8)\, [\log (\Lambda^{2}/B) 
+ F^{ba}(\gamma'^{\rm ren},d^{\rm ren}, \cdots)] $,
where $F^{ba}$ denote finite corrections.


\begin{figure}[tbp]
\includegraphics[scale=0.3]{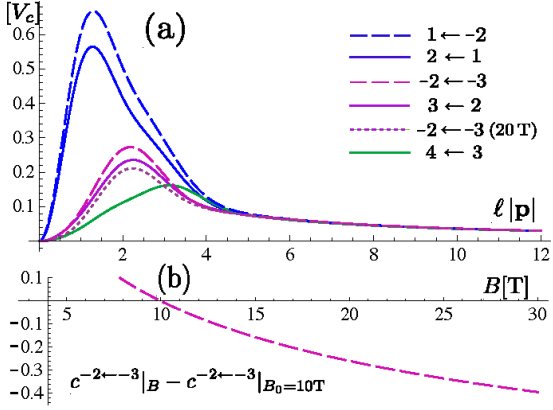}  
\caption{
(a)~Momentum profiles of the many-body corrections
$\triangle \epsilon^{b\leftarrow a}/\eta^{b\leftarrow a}$ for some typical channels,
with $v= 1.1 \times 10^{6}$m/s,  $\gamma_{1} =404\, {\rm meV}$, $\triangle = 18\, {\rm meV}$ 
and $v_{4}/v= -0.04$ at $B$= 10\,T.
(b)~$c^{-2 \leftarrow -3}|_{B} -c^{-2 \leftarrow -3}|_{B_{0}}$ reveals
the running of $v^{\rm ren}|_{B}$. 
}
\end{figure}


Figure 3~(a) shows for some typical resonance channels  
the (rescaled) momentum profiles $\gamma_{\bf p}^{2}\, [\cdots]$ 
in $\triangle \epsilon^{b\leftarrow a}$ of Eq.~(\ref{dEba}),
which, when integrated over $\ell |{\bf p}|$, give
$\triangle \epsilon^{b\leftarrow a}/\eta^{b\leftarrow a}$
in units of $V_{c}$.
Note that the slowly decreasing high-momentum tail
$\sim (\sqrt{2}/4)/(\ell\, |{\bf p}|)$, 
common to all profiles, is responsible for the UV divergence.
This numerically verifies the UV scaling law~(\ref{dEscale}) of 
the ratios $\triangle \epsilon^{b\leftarrow a}/\eta^{b\leftarrow a}$.

For renormalization let us refer to a specific channel $(b_{0}\leftarrow a_{0})$
and choose to define $v^{\rm ren}$ and other renormalized parameters by writing 
$\epsilon_{\rm exc}^{b_{0} \leftarrow a_{0}} 
= \omega_{c}^{\rm ren}\, (\eta_{b_{0}}- \eta_{a_{0}})$ at magnetic field $B$,
or equivalently, 
$ (\triangle \epsilon^{b_{0}a_{0}})^{\rm ren}=0$, 
which yields $\delta v = - (\alpha/\sqrt{2}\, \epsilon_{b})\, c^{b_{0}\leftarrow a_{0}}$.
The renormalized velocity then runs with $B$, 
\begin{equation}
v^{\rm ren}|_{B} = 
v^{\rm ren}|_{B_{0}} 
+ {\alpha\over{\sqrt{2}\, \epsilon_{b}}}\,
\Big\{c^{b_{0}\leftarrow a_{0}}|_{B} -c^{b_{0}\leftarrow a_{0}}|_{B_{0}} \Big\}, 
\label{runningvB}
\end{equation}
and decreases gradually with increasing $B$. 
The leading correction
$c^{b_{0}\leftarrow a_{0}}|_{B} -c^{b_{0}\leftarrow a_{0}}|_{B_{0}} 
\sim  -(\sqrt{2}/8)\, \log (B/B_{0})$ is logarithmic but corrections
coming from finite terms $F^{b_{0}a_{0}}$ are equally important for relatively low magnetic fields.
For definiteness let us take $L_{-2} \leftarrow L_{-3}$ as the reference channel, 
as chosen experimentally.~\cite{HJTS}
For this channel the contribution from the low-momentum region decreases with $B$, 
as seen from the $(-2\leftarrow -3)$ profiles for $B=(10{\rm T}, 20{\rm T})$ 
in Fig.~3~(a), and numerically  the correction is roughly doubled,~\cite{fncba}
\begin{equation}
c^{-2\leftarrow -3}|_{B} -c^{-2 \leftarrow -3}|_{B_{0}} 
\approx  -2.1 \times {\sqrt{2}\over{8}}\, \log {B\over{B_{0}}}
\end{equation} 
over the range $10\, {\rm T} \lesssim B \lesssim 30\, {\rm T}$,
as shown in Fig.~3~(b).
One can multiply it by factor 
$\alpha/(\sqrt{2}\, \epsilon_{b}v) \sim 1.5/\epsilon_{b}\sim 0.3$
(with $\epsilon_{b} \sim 5$)
to estimate the rate of decrease in $v^{\rm ren}|_{B}$ with $B$, 
which is about 10\% for $B=10{\rm T} \rightarrow 20 {\rm T}$.

The renormalized Coulombic corrections in all other channels are thereby fixed 
uniquely,
\begin{equation}
(\triangle \epsilon^{ba})^{\rm ren} 
= V_{c}\, \eta^{b\leftarrow a}\, (c^{b\leftarrow a}
- c^{b_{0}\leftarrow a_{0}})|_{B} . 
\end{equation} 
These observable corrections are essentially calculated 
from the profiles in the low-momentum region $\ell\, |{\bf p}| \lesssim 15$.

It is enlightening to write the resonance energies as 
\begin{eqnarray}
\epsilon_{\rm exc}^{b \leftarrow a} &=&
(\eta_{b}- \eta_{a})\,  
\Big[ \omega_{c}^{\rm ren} + V_{c}\, \tilde{\triangle}\, c^{b, a} \Big],
\label{Eexcscale} \\
 \tilde{\triangle}c^{b, a} &\equiv&  {\eta^{b\leftarrow a} \over{\eta_{b}- \eta_{a}}}\,
(c^{b\leftarrow a} - c^{b_{0}\leftarrow a_{0}})|_{B},
\end{eqnarray}
so that the Coulombic corrections $V_{c}\, \tilde{\triangle}\, c^{b, a}$ 
seemingly arise relative to $\omega_{c}^{\rm ren}$.
Using the set of parameters in Eq.~(\ref{expparameters}), one finds, 
for some typical intraband channels, 
\begin{eqnarray}
\tilde{\triangle} c^{5,4}
&\stackrel{\nu=16}{=}& -0.317,\ \ (-0.259), \nonumber\\
\tilde{\triangle} c^{4, 3}
&\stackrel{\nu=12}{=}&   -0.223,\ \ (-0.175),  \nonumber\\
\tilde{\triangle} c^{3,2}
&\stackrel{\nu=8}{=}&   -0.073,\ \ (-0.040),  \nonumber\\
\tilde{\triangle} c^{2,1}
&\stackrel{\nu=4}{=}&   0.650,\ \ (0.533),  \nonumber\\
\tilde{\triangle} c^{1,-2}
&\stackrel{\nu=-4}{=}&   0.897,\ \ (0.675),  \nonumber\\
\tilde{\triangle} c^{-2,-3}
&\stackrel{\nu=-8}{=}&  0,\ \ (0),  \nonumber\\
\tilde{\triangle} c^{-3, -4}
&\stackrel{\nu=-12}{=}&   -0.165,\ \ (-0.150),  \nonumber\\
\tilde{\triangle} c^{-4,-5}
&\stackrel{\nu=-16}{=}& -0.275,\ \ (-0.247), 
\label{BLinter}
\end{eqnarray}
at $B=10$T  ($B=20$T).
For comparison, setting $v_{4}=\triangle=0$ yields 
the electron-hole symmetric values  
$\tilde{\triangle} c^{1,-2}=0.815$, $\tilde{\triangle} c^{-2,-3}=0$, 
$\tilde{\triangle} c^{-3,-4} =-0.159$ and
$\tilde{\triangle} c^{-5,-4}= -0.303$ at $B=10$T.
Similarly, some interband channels yield
\begin{eqnarray}
\tilde{\triangle} c^{3,-2}
&\stackrel{\nu=-4}{=}& 0.353  ,\ \ (0.375),  \nonumber\\
\tilde{\triangle} c^{2,-3}
&\stackrel{\nu=-8}{=}& 0.370  ,\ \ (0.363),  \nonumber\\
\tilde{\triangle} c^{3,-4}
&\stackrel{\nu=-12}{=}& 0.261  ,\ \ (0.278).
\end{eqnarray}
Note that those corrections are  ordered regularly in magnitude 
for a sequence of resonances;
see also the theoretical curves for $\epsilon_{\rm exc}^{b \leftarrow a}$
in Figs.~4(b) and 4(d).

\begin{figure}[tbp]
\includegraphics[scale=0.33]{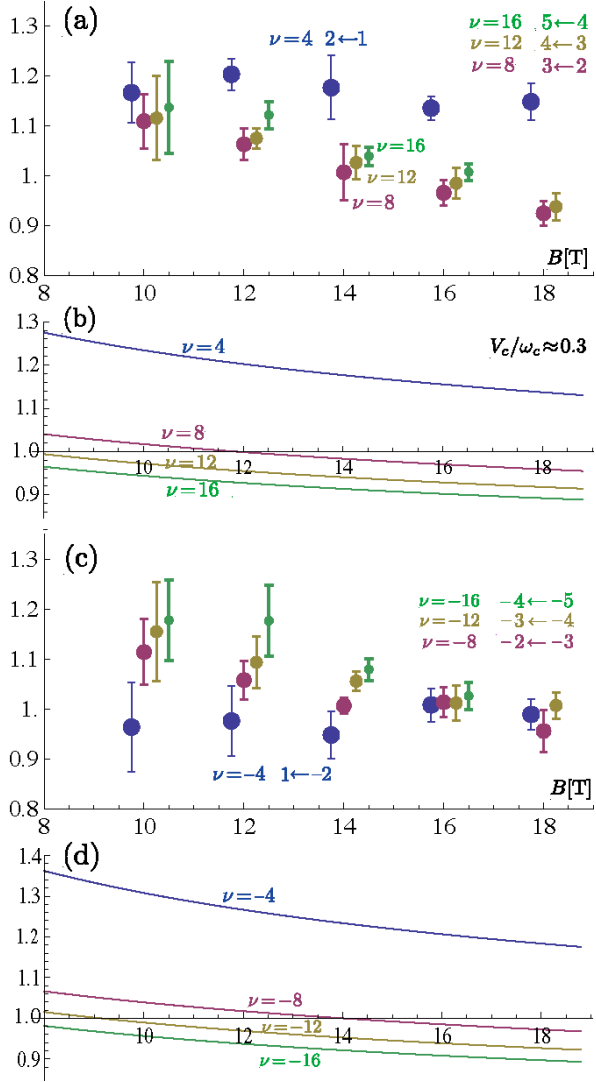}  
\caption{
Cyclotron-resonance  data of Ref.~\onlinecite{HJTS} with error bars, reorganized in the form  
$\epsilon^{b\leftarrow a}_{\rm exc}/(\eta^{b} - \eta^{a})$
and plotted in units of $\omega_{c}=\sqrt{2}\, v_{0}/\ell$
(with $v_{0} = 1.1 \times 10^{6}$m/s) for 
$\gamma_{1} \approx 404\, {\rm meV}$,
$\triangle \approx  18\, {\rm meV}$ and 
$v_{4}/v \equiv - \gamma_{4}/\gamma_{0} \approx - 0.04$.
(a)~Electron data; for clarity the data points, originally 
at $B$=(10, 12, 14, 16, 18)\,T, are slightly shifted in $B$.
(b)~Theoretical expectation according 
to Eq.~\~(\ref{runningvB}), with  
$V_{c}/\omega_{c}\approx 0.3$ (or $\epsilon_{b}\approx 5)$.
(c)~Hole data.
(d)~Theoretical curves.
}
\end{figure}


Our formula~(\ref{Eexcscale}) summarizes the effect of renormalization 
in a concise form and is also useful in analyzing the experimental results.
One may rescale the observed excitation energies 
$\epsilon^{b\leftarrow a}_{\rm exc}$
in the form $\epsilon^{b\leftarrow a}_{\rm exc}/(\eta_{b}- \eta_{a})$
and plot them in units of $1/\ell \propto \sqrt{B}$ for each given value of $B$. 
The Coulombic many-body effect will then be
seen as a variation in characteristic velocity $v^{\rm ren}|_{B}[1 + O(V_{c})]$
from one resonance to another, and 
a deviation of $\omega_{c}^{\rm ren}$ from $\sqrt{B}$ behavior
would indicate the running of $v^{\rm ren}$ with $B$.

Figures 4(a) and 4(c) show such plots for the series of cyclotron resonance 
reported in Ref.~\onlinecite{HJTS}.  Both electron and hole data are plotted  
in units of $\omega_{c} = \sqrt{2}\, v_{0}/\ell \approx 40 \sqrt{B[{\rm T}]}$ meV 
(with reference velocity $v_{0}= 1.1 \times 10^{6}$ m/s kept fixed).
These plots offer a closer look into the plots in Fig.~2.

They are  to be compared with Figs.~4(b) and 4(d), which illustrate 
how each resonance would behave with $B$, according to Eq.~(\ref{Eexcscale}), 
for $V_{c}\approx 11 \sqrt{B[{\rm T}]}$ meV (or $\epsilon_{b}\approx 5)$.
The $\nu=-8$ curve for $\epsilon_{\rm exc}^{-2 \leftarrow -3}|_{\nu=-8}$,
in particular, 
represents the running of $v^{\rm ren}|_{B}$ 
according to Eq.~(\ref{runningvB}) (normalized to 1 at $B_{0}=14$T).
These theoretical curves and experimental data look similar 
but differ in details.
They are not quite consistent, but there are some notable features:
(i) The $2\leftarrow 1$ and $1\leftarrow -2$ resonances (the $\nu=\pm 4$ data)
appear distinct from the rest, especially in their variation with $B$. 
In addition, the $\nu=\pm 4$ theoretical curves are separated 
from the rest by appreciable Coulombic gaps,
but such a gap is not seen in the hole data.
This would indicate, as noted in Sec.~II, that 
the $n=(0, 1)$ sector in bilayer graphene is significantly 
modified from the naive one we have supposed.

(ii) The $\nu= (8,12,16)$ electron data and the $\nu=(-8,-12,-16)$ hole data 
show a general trend to decrease with $B$, consistent with possible running 
of $v^{\rm ren}$ with $B$.
Such ($\sim$ logarithmic) running of $v^{\rm ren}|_{B}$ 
is a direct consequence of renormalization
and is thus the key signature of the Coulomb interaction.
In both electron and hole data $v^{\rm ren}$ appears to run in the same way 
at a rate somewhat faster than naively expected. 
Such enhanced running could in part be attributed to 
possible quantum screening~\cite{KSpzm} of the Coulomb interaction in graphene 
such that $\epsilon_{b}$ is effectively larger~\cite{KSbgr} for lower $B$.



\begin{figure}[tbp]
\includegraphics[scale=0.9]{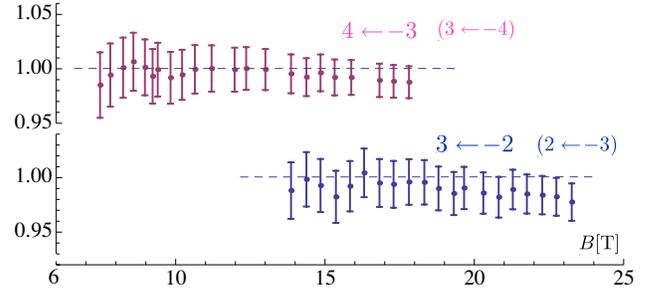}  
\caption{Interband cyclotron resonances in bilayer inclusions 
in multilayer epitaxial graphene
reported by Orlita {\it et al}. 
Here some data from Ref.~\onlinecite{OFB} are analyzed in the same way as 
in Fig.~4, using $v_{0} = 1.02 \times 10^{6}$m/s and
$\gamma_{1} \approx 385\, {\rm meV}$ 
for zero asymmetry $\triangle= \gamma_{4}=0$.
}
\end{figure}

Interband cyclotron resonance was recently observed by Orlita {\it et al.}~\cite{OFB}
in bilayer inclusions in multilayer epitaxial graphene on the C-face of SiC.
They identify some $n-1 \leftarrow - n$ [or $n \leftarrow - (n-1)$] resonances
with $n\ge 3$ and obtain, via fitting, 
$v \approx 1.02\times 10^6$ m/s  and $\gamma_{1}\approx 385$ meV, 
which are somewhat smaller than those for bilayer graphene. 

Some of their data are analyzed according to our formula~(\ref{Eexcscale})  in Fig.~5;
there we have set $\gamma_{4}=\triangle=0$ since this experiment searched 
for no intraband resonances which would clarify a possible electron-hole asymmetry. 
The data appear to indicate slight running of $v^{\rm ren}$ with $B$, 
far slower than in the data in Fig.~4 on bilayer graphene. 
This suggests that the Coulomb interaction could be significantly weaker 
(or more efficiently screened) in multilayered epitaxial graphene 
than in exfoliated bilayer graphene.

\section{summary and discussion}

Experiment suggests that bilayer graphene has intrinsic electron-hole asymmetry
due to subleading intralayer and interlayer couplings.
In this paper we have studied cyclotron resonance in bilayer graphene 
with such asymmetry taken into account.

The set of asymmetry parameters, $\triangle \approx 18$ meV and 
$\gamma_{4}/\gamma_{0} \approx 0.04$
derived from independent measurements, 
entails a considerable modification of Landau levels in bilayer graphene
and improves the theoretical fit to the data on cyclotron resonance 
between higher levels $|n|\ge 2$ in both electron and hole bands.
In contrast, the fit to the data 
on $(2\leftarrow 1)$ and $(1\leftarrow -2)$ resonances
appears somewhat puzzling, and this suggests 
that the zero- and pseudo-zero-mode Landau levels $n= (0,\pm1)$ 
are further affected by some sources
other than $\triangle$ and  $\gamma_{4}$.
It would be important to clarify, both theoretically and experimentally, 
the detailed structure of this special sector in bilayer graphene.

The Coulombic many-body corrections to cyclotron resonance in graphene,
unlike in standard quantum-Hall systems,
are afflicted with UV divergences, and
one has to carry out renormalization to extract genuine observable corrections.
We have shown how to perform renormalization for bilayer graphene 
under a magnetic field 
by first constructing necessary conterterms in free space.
This renormalization program, formulated analytically, can equally be handled
numerically in practical calculations by use of the reduced matrix Hamiltonian 
$\hat{\cal H}_{\rm red}^{\rm ren}$ in Eq.~(\ref{reducedH})
and counterterm
$\delta_{\rm ct}\hat{\cal H}_{\rm red}^{\rm ren}$ in Eq.~(\ref{matrixct}).
As a further illustration, we present the renormalization program 
for monolayer graphene with a possible valley gap in Appendix B.

Equation~(\ref{Eexcscale}) summarizes the effect of renormalization
on cyclotron-resonance energies in a neat and concise form. 
This formula is also useful in analyzing the experimental data; 
it magnifies possible effects of the Coulombic corrections per channel 
and running of the renormalized velocity $v^{\rm ren}$ with $B$,
as we have seen in Sec.~III.  
In particular, the nearly logarithmic running of $v^{\rm ren}$ is 
a direct consequence of renormalization, specific to graphene.~\cite{EGM,AddRef}
It is remarkable that such a renormalization effect is apparently seen in the data.

More detailed measurements of cyclotron resonance, both intraband and interband ones, 
are highly desired to pin down the many-body effects
as well as the structure of the zero-mode and pseudo-zero-mode sector in bilayer graphene.

\acknowledgments

This work was supported in part by a Grant-in-Aid for Scientific Research
from the Ministry of Education, Science, Sports and Culture of Japan 
(Grant No. 21540265).

\appendix

\section{Analysis of divergence}

In this appendix we examine the UV structure 
of the Coulomb exchange correction.
Let us first look at ${\cal H}_{+}$ in Eq.~(\ref{Hbilayer}) 
and construct the electron propagator in free space,
\begin{equation}
\langle \Psi (x) \Psi^{\dag}(x') \rangle
= \langle x| i/(i\partial_{t} -{\cal H}_{+}) |x' \rangle
\end{equation}
with $|x\rangle \equiv|{\bf x}\rangle\, | t \rangle$.
We divide ${\cal H}_{+}$ into $2\times 2$ blocks,
\begin{eqnarray}
{\cal H}_{+}&=& \left(
\begin{array}{cc}
m\, \sigma_{3}&    v\,P+ v_{4}\,Q \\
 v\, P+v_{4}\,Q^{\dag}\ 
& - m\, \sigma_{3} +\triangle 
+ \gamma_{1}\,\sigma_{1}   \\
\end{array}
\right) ,
\label{HplusAp}
\end{eqnarray}
where we have set  $m\equiv u/2$; 
$P= p^{\dag} \sigma_{+}
+p\, \sigma_{-}$  with 
 $p=p_{x}+i\, p_{y}$ and 
$\sigma_{\pm}= (\sigma_{1}\pm i \sigma_{2})/2$;
$Q={\rm diag.}(p^{\dag}, p)$;
the unit matrix $1$, which, e.g., 
multiplies $\triangle$ in Eq.$~(\ref{HplusAp})$, 
will be suppressed in what follows.

We go to the Fourier $({\bf p},\omega)$ space 
and invert $(\omega -{\cal H}_{+} )$
in this $2\times 2$ block form. 
A direct calculation yields
$\langle \Psi \Psi^{\dag}\rangle_{jk} = i\, N_{jk}/D$ with
\begin{eqnarray}
N_{11} 
&=& \Gamma + \{ (v^{2}+ v_{4}^{2})\, \gamma_{1} 
+ 2v_{4} v\, (\omega - \triangle) \}\, P\sigma_{1} P 
\nonumber\\
&& + m\, (\Xi -2\, \triangle\, \omega + \triangle^{2} -\gamma_{1}^{2} )\, \sigma_{3},
\nonumber\\
N_{22} &=& \Gamma
- 2\, v\, ( v\,\triangle - v_{4}\gamma_{1})\, {\bf p}^{2} 
\nonumber\\
&&+ \omega\, (\gamma_{1}^{2}- \triangle^{2})
+ \triangle \, ( \Xi + 2m^{2})  \nonumber\\
&& -m\, \Xi\, \sigma_{3} 
+ \{\gamma_{1}\, (\omega^{2} -m^{2}) 
+ 2\, v_{4}v\, {\bf p}^{2}\, \omega \}\, \sigma_{1},
 \nonumber\\
N_{12} 
&=& v\, (\omega^{2} - v_{+}v_{-}\,  {\bf p}^{2} + m^{2})\, P \nonumber\\
&& + 2m\, v_{4}\, \sigma_{3} P 
+ v_{4}\, \Xi\, Q\nonumber\\
&&+(v_{4}\, \gamma_{1} -v\, \triangle)\,( \omega + m\,\sigma_{3})\, P 
\nonumber\\
&&+(v\, \gamma_{1} - v_{4} \triangle )\,( \omega + m\,\sigma_{3})\, Q,
\end{eqnarray}
and $N_{21}= (N_{12})^{\dag}$, where
\begin{eqnarray}
\Gamma
&=&\omega\, \{ (\omega -\triangle)^{2}
- (v^{2}+ v_{4}^{2})\, {\bf p}^{2} - m^{2} -\gamma_{1}^{2}\} 
\nonumber\\
&&+ \{ (v^{2}+v_{4}^{2})\, \triangle - 2\, v_{4}v\, \gamma_{1}\}\, {\bf p}^{2},  
\nonumber\\
\Xi &=& \omega^{2} +v_{+}v_{-}\,  {\bf p}^{2} - m^{2};
\end{eqnarray}
$v_{\pm}\equiv v \pm v_{4}$ and 
$P\sigma_{1} P = (p^{\dag})^{2}\, \sigma_{+} + p^{2}\, \sigma_{-}$.
The denominator $D= {\rm det}\, (\omega -{\cal H}_{+})$ is cast in the form
\begin{eqnarray}
D&=& \{ \omega^{2} - \triangle\, \omega 
- (v^{2}+ v_{4}^{2})\, {\bf p}^{2} - m^{2} \}^{2}
\nonumber\\
&&- (\gamma_{1}\, \omega + 2 v_{4}v\, {\bf p}^{2})^{2} \nonumber\\
&& - m^{2}\, ( 4 \, v^{2} {\bf p}^{2} - \gamma_{1}^{2} - \triangle^{2}),
\label{Denom}
\end{eqnarray}
which, for $u=2m\rightarrow 0$, leads to the band spectra $\{\epsilon_{i}\}$ 
in Eq.~(\ref{Eip}).

The Coulomb exchange correction to $O(\alpha/\ell)$ is written as
the convolution  $\sum_{\bf k}v_{\bf k}\, iS({\bf p+k})$ of 
$v_{\bf k}=2\pi\alpha/|{\bf k}|$ and the instantaneous limit of the electron propagator 
$\langle \Psi\, \Psi^{\dag}\rangle|_{t'=t} 
= \int (d\omega/2\pi)\, \langle \Psi\, \Psi^{\dag}\rangle_{\omega, {\bf p}} = i S({\bf p})$.
In particular, divergences arise from the portion of $S({\bf p})$, that decreases
like $1/|{\bf p}|$ or slower for ${\bf p} \rightarrow  \infty$, and we shall focus on that portion.

Integration over $\omega$, with the standard boundary condition, 
is readily carried out, yielding, e.g., 
\begin{equation}
\int {d \omega\over{2\pi}} {1 \over{D(\omega)}}
= i\, {a+b+c+d\over{(a+c)(b+c)(a+d)(b+d)}},  
\end{equation} 
where $D(\omega) = (\omega - a)(\omega - b)(\omega + c)(\omega + d)$ $(a, b, c, d>0)$
is short for $D$ in Eq.~(\ref{Denom}).
When the integrand is $\omega/D$, replace the numerator on the the right-hand side with 
$ab - cd$; for $\omega^{2}/D$ replace it with $-ab(c+d) - (a+b) cd$.
Actually, this structure combined with the band spectra~(\ref{Eip}) gives rise to a simple rule 
to handle the integration over $\omega$: One can effectively replace
\begin{eqnarray}
\omega &\rightarrow&  (\triangle - r\, \gamma_{1})/2 + \cdots, \nonumber\\
\omega^{2} &\rightarrow&  - v_{+}v_{-}\, {\bf p}^{2} + \cdots,
\label{omegaintegral}
\end{eqnarray}  
in $N_{jk}$ in evaluating their large-${\bf p}$ behavior.

One may further note the following:
(i)~The denominator $(a+c)...(b+d) \approx 16v^{2}v_{+}v_{-}\, ({\bf p}^{2})^{2}$
for large ${\bf p}$.
(ii)~For $v_{4}\rightarrow 0$ and $\triangle\rightarrow 0$, $D(\omega)$ becomes 
an even function of $\omega$.  As a result, odd powers of $\omega$ in $N_{jk}$ 
necessarily lead to $v_{4}\gamma_{1}$ or $\triangle$, and are one power of $|{\bf p}|$ 
less than naively expected; e.g., 
$ab-cd \approx 2(v\triangle  - v_{4}\, \gamma_{1})\, |{\bf p}|$; 
corrections of the form $v_{4}\, m$ do not arise since $D$ is even in $m= u/2$.

With this in mind one can now retain only the portion 
\begin{eqnarray}
N_{11} 
&\approx & \Gamma + \{ (v^{2}+ v_{4}^{2})\, \gamma_{1} 
+ 2v_{4} v\, (\omega - \triangle) \}\, P\sigma_{1}P,
\nonumber\\
N_{22} 
&\approx & \Gamma - 2\, v\, ( v\,\triangle - v_{4}\gamma_{1})\, {\bf p}^{2} 
+ (\gamma_{1}\, \omega^{2} + 2\, v_{4}v\, {\bf p}^{2}\, \omega )\, \sigma_{1},
\nonumber\\
N_{12}&\approx & v\, (\omega^{2} -v_{+}v_{-}\, {\bf p}^{2} )\, P,
\label{Nreduced}
\end{eqnarray}
for further consideration.
Note that, in view of Eq.~(\ref{omegaintegral}), the particular combination $\Xi$,
despite its appearance,  yields no divergent correction. 
As a result, terms $\propto \sigma_{3}$ in $N_{11}$ and  $N_{22}$
lead to finite  corrections.  
This fact has the important consequence that the interlayer bias $u=2m$ 
requires no infinite renormalization.

The structure of the $O(\alpha/\ell)$ Coulombic selfenergy correction  
precisely reflects the structure of $S({\bf p})$ 
after convolution with $v_{\bf p}\propto 1/|{\bf p}|$.
Let us therefore compare Eq.~(\ref{Nreduced}) with ${\cal H}_{+}$ in Eq.~(\ref{HplusAp}).
Applying first the rule~(\ref{omegaintegral}) to $N_{12}$ in Eq.~(\ref{Nreduced}) yields 
the asymptotic form
\begin{equation}
\langle \Psi \Psi^{\dag}\rangle_{12}|_{t=t'} 
=  P/(2\, |{\bf p}|) + \cdots.
\end{equation} 
This $P/2\, |{\bf p}|$ term leads to a divergent correction of the form 
$\propto P\, \log \Lambda^{2}$, which 
requires renormalization of velocity~\cite{velrenorm} $v$. 
Actually this leading form $P/2\, |{\bf p}|$ is the same as 
the one obtained previously~\cite{KScr}
for $v_{4}= \triangle =0$, and this implies 
that (the divergent part of) velocity renormalization is unaffected 
by the electron-hole asymmetry $\propto v_{4}, \triangle$.
On the other hand, the portion $\propto Q$ in $\langle \Psi \Psi^{\dag}\rangle_{12}$
has the asymptotic form $\propto Q/{\bf p}^{2}$ which leads to no divergence 
and this means that $v_{4}$ remains finite.

Let us next consider $\langle \Psi \Psi^{\dag}\rangle_{11}$ or $N_{11}$ in Eq.~(\ref{Nreduced}).
Its $P\sigma_{1}P$ portion has the asymptotic structure   $P\sigma_{1}P/|{\bf p}|^{3}$,
which, though potentially singular, actually yields no divergent 
correction via symmetric ${\bf p}$ integration
[since  $P\sigma_{1}P \propto p^{2}$ or $(p^{\dag})^{2}$].
The remaining portion $\propto \Gamma$ is common to $N_{11}$ and $N_{22}$.  
Those common $\Gamma$ terms, though leading to a divergent correction, 
simply shift the zero of energy and are of no physical relevance.
One can now eliminate this $\Gamma$ term 
from $\langle \Psi \Psi^{\dag}\rangle_{22}|_{t=t'}$
and determine its asymptotic form 
\begin{equation}
\langle \Psi \Psi^{\dag}\rangle_{22}|_{t=t'} 
= {v\, \gamma_{1} -  v_{4}\, \triangle \over{4\, v_{+}v_{-}|{\bf p}|}}\, 
\sigma_{1}  + {v\, \triangle - v_{4}\, \gamma_{1} \over{2\, v_{+}v_{-} |{\bf p}|}} + \cdots.
\end{equation} 
This implies that both $\gamma_{1}$ and $\triangle$ undergo infinite renormalization.
Evaluating the convolution integral $\int d^{2}{\bf k}\, v_{\bf k}\, S({\bf p+k})$
with momentum cutoff $\Lambda$ eventually leads to the counterterms 
in Eqs.~(\ref{deltav})~$\sim$~(\ref{deltagamma}).

\section{monolayer graphene with a valley gap}
In this appendix we outline the renormalization prescription for monolayer graphene 
with a possible valley gap $M$. The effective Hamiltonian is written as
\begin{equation}
{\cal H}_{+} =  v\, {\bf p} \cdot \vec{\sigma} + M\, \sigma_{3}
\end{equation}
at one valley and acts on a two-component spinor of the form 
$\Psi =(\psi_{A}, \psi_{B})^{\rm t}$.
One can pass to another valley by setting $M\rightarrow -M$ and 
$\Psi\rightarrow \Psi' = (-\psi'_{B}, \psi'_{A})^{\rm t}$.

Using the instantaneous propagator 
\begin{eqnarray}
\langle \Psi \Psi^{\dag}\rangle_{t=t'} 
= {1\over{2}}\, { v\, {\bf p} \cdot \vec{\sigma} 
+ M\, \sigma_{3} \over{\sqrt{v^{2}\, {\bf p}^{2} + M^{2}}}} 
\end{eqnarray}
one can calculate the Coulombic selfenergy correction to $O(\alpha)$
and find divergences of the form
\begin{equation}
\delta v \sim 
 -(\alpha/8 \epsilon_{b})\, \log \Lambda^{2},\
\delta M = 2\,  (M^{\rm ren}/v^{\rm ren})\, \delta v,
\end{equation}
which implies that the mass gap $M$, as well as $v$, undergoes renormalization.

Let us now pass to the $B\not =0$ case and denote the zeroth order Landau-level
spectrum as $\epsilon_{n}^{\rm ren} = \omega_{c}^{\rm ren}\, \eta_{n}$ with 
$\omega_{c}^{\rm ren} = \sqrt{2}\, v^{\rm ren}/\ell$ and 
\begin{equation}
\eta_{n} = {\rm sgn}[n]\, \sqrt{ |n|+ (M^{\rm ren}/\omega_{c}^{\rm ren})^{2}}.
\label{etaML}
\end{equation}
Letting $\delta_{\rm ct}= \delta v\, \partial/\partial v^{\rm ren} + 
\delta M\, \partial/\partial M^{\rm ren}$
act on $\epsilon_{n}^{\rm ren}$  
then yields the counterterm  $\delta_{\rm ct}\epsilon_{n}^{\rm ren}$.
In particular, one finds that ${\cal D}\eta_{n}$ in Eq.~(\ref{etaba}) 
is now replaced by
\begin{equation}
{\cal D}\eta_{n}\rightarrow   (M^{\rm ren}/\omega_{c}^{\rm ren})^{2}/\eta_{n}. 
\label{Detan}
\end{equation}
With Eqs.~(\ref{etaML}) and (\ref{Detan}), 
velocity and mass renormalization for monolayer graphene in a magnetic field
is carried out according to formula~(\ref{Eexcscale}); 
we have checked numerically that this renormalization program works correctly.



\begin{thebibliography}{99}
\bibitem{fndH} For an eigenmode $|\psi_{n}\rangle$ of $H$ with eigenvalue $\epsilon_{n}$
one generally finds that 
$\langle \psi_{n}| \delta H | \psi_{n}\rangle = \delta \epsilon_{n}$ for an arbitrary variation $\delta$.
\end{thebibliography}
\end{document}